\documentclass[%
prl,
twocolumn,
superscriptaddress,
 amsmath,amssymb,
 aps,
floatfix,
]{revtex4-2}

\usepackage{graphicx}
\usepackage{dcolumn}
\usepackage{bm}
\usepackage[mathlines]{lineno}
\usepackage{upgreek}  
\usepackage{multirow}  
\usepackage{xcolor}
\usepackage{placeins}           
\usepackage[colorlinks=true,citecolor=blue,filecolor=blue,linkcolor=blue,urlcolor=blue]{hyperref}

\newcommand{\bologna}{\affiliation{Department of Physics and Astronomy, University of Bologna and INFN-Bologna, 40126 Bologna, Italy}}
\newcommand{\chicago}{\affiliation{Department of Physics \& Kavli Institute for Cosmological Physics, University of Chicago, Chicago, IL 60637, USA}}
\newcommand{\coimbra}{\affiliation{LIBPhys, Department of Physics, University of Coimbra, 3004-516 Coimbra, Portugal}}
\newcommand{\columbia}{\affiliation{Physics Department, Columbia University, New York, NY 10027, USA}}
\newcommand{\lngs}{\affiliation{INFN-Laboratori Nazionali del Gran Sasso and Gran Sasso Science Institute, 67100 L'Aquila, Italy}}
\newcommand{\mainz}{\affiliation{Institut f\"ur Physik \& Exzellenzcluster PRISMA$^{+}$, Johannes Gutenberg-Universit\"at Mainz, 55099 Mainz, Germany}}
\newcommand{\heidelberg}{\affiliation{Max-Planck-Institut f\"ur Kernphysik, 69117 Heidelberg, Germany}}
\newcommand{\munster}{\affiliation{Institut f\"ur Kernphysik, Westf\"alische Wilhelms-Universit\"at M\"unster, 48149 M\"unster, Germany}}
\newcommand{\nikhef}{\affiliation{Nikhef and the University of Amsterdam, Science Park, 1098XG Amsterdam, Netherlands}}
\newcommand{\nyuad}{\affiliation{New York University Abu Dhabi - Center for Astro, Particle and Planetary Physics, Abu Dhabi, United Arab Emirates}}
\newcommand{\purdue}{\affiliation{Department of Physics and Astronomy, Purdue University, West Lafayette, IN 47907, USA}}
\newcommand{\rice}{\affiliation{Department of Physics and Astronomy, Rice University, Houston, TX 77005, USA}}
\newcommand{\stockholm}{\affiliation{Oskar Klein Centre, Department of Physics, Stockholm University, AlbaNova, Stockholm SE-10691, Sweden}}
\newcommand{\subatech}{\affiliation{SUBATECH, IMT Atlantique, CNRS/IN2P3, Nantes Universit\'e, Nantes 44307, France}}
\newcommand{\torino}{\affiliation{INAF-Astrophysical Observatory of Torino, Department of Physics, University  of  Torino and  INFN-Torino,  10125  Torino,  Italy}}
\newcommand{\ucsd}{\affiliation{Department of Physics, University of California San Diego, La Jolla, CA 92093, USA}}
\newcommand{\wis}{\affiliation{Department of Particle Physics and Astrophysics, Weizmann Institute of Science, Rehovot 7610001, Israel}}
\newcommand{\zurich}{\affiliation{Physik-Institut, University of Z\"urich, 8057  Z\"urich, Switzerland}}
\newcommand{\paris}{\affiliation{LPNHE, Sorbonne Universit\'{e}, CNRS/IN2P3, 75005 Paris, France}}
\newcommand{\freiburg}{\affiliation{Physikalisches Institut, Universit\"at Freiburg, 79104 Freiburg, Germany}}
\newcommand{\napels}{\affiliation{Department of Physics ``Ettore Pancini'', University of Napoli and INFN-Napoli, 80126 Napoli, Italy}}
\newcommand{\nagoya}{\affiliation{Kobayashi-Maskawa Institute for the Origin of Particles and the Universe, and Institute for Space-Earth Environmental Research, Nagoya University, Furo-cho, Chikusa-ku, Nagoya, Aichi 464-8602, Japan}}
\newcommand{\laquila}{\affiliation{Department of Physics and Chemistry, University of L'Aquila, 67100 L'Aquila, Italy}}
\newcommand{\tokyo}{\affiliation{Kamioka Observatory, Institute for Cosmic Ray Research, and Kavli Institute for the Physics and Mathematics of the Universe (WPI), University of Tokyo, Higashi-Mozumi, Kamioka, Hida, Gifu 506-1205, Japan}}
\newcommand{\kobe}{\affiliation{Department of Physics, Kobe University, Kobe, Hyogo 657-8501, Japan}}

\newcommand{\kit}{\affiliation{Institute for Astroparticle Physics, Karlsruhe Institute of Technology, 76021 Karlsruhe, Germany}}
\newcommand{\tsinghua}{\affiliation{Department of Physics \& Center for High Energy Physics, Tsinghua University, Beijing 100084, China}}

\begin{document}

\title{Search for New Physics in Electronic Recoil Data from XENONnT}

\author{E.~Aprile}\columbia
\author{K.~Abe}\tokyo
\author{F.~Agostini}\bologna
\author{S.~Ahmed Maouloud}\paris
\author{L.~Althueser}\munster
\author{B.~Andrieu}\paris
\author{E.~Angelino}\torino
\author{J.~R.~Angevaare}\nikhef
\author{V.~C.~Antochi}\stockholm
\author{D.~Ant\'on Martin}\chicago
% ...

\collaboration{XENON Collaboration}
\email[]{xenon@lngs.infn.it}
\noaffiliation

\date{\today}

\begin{abstract}
We report on a blinded analysis of low-energy electronic-recoil data from the first science run of the XENONnT dark matter experiment. Novel subsystems and the increased 5.9\,tonne liquid xenon target reduced the background in the (1, 30)\,keV search region to $(15.8 \pm 1.3)$\,events/(tonne$\times$year$\times$keV), the lowest ever achieved in a dark matter detector and $\sim$5 times lower than in XENON1T. With an exposure of 1.16\,tonne-years, we observe no excess above background and set stringent new limits on solar axions, an enhanced neutrino magnetic moment, and bosonic dark matter.
\end{abstract}

\maketitle

In 2020 we reported an unexpected excess of electronic recoil\,(ER) events below $\sim$7\,keV in the XENON1T dark matter\,(DM) experiment~\cite{XENON:2020rca}. The excess was compatible with decays from trace amounts of tritium, the presence of which we were unable to confirm or exclude at the time. The result was also interpreted as physics beyond the Standard Model\,(BSM) such as solar axions, bosonic DM with a mass of $\sim$2.3\,keV/c$^2$, solar neutrinos with enhanced magnetic moment, and many other models~\cite{Leane:2022bfm}. ER data are also used to search for fermionic dark matter, as recently reported by PandaX-4T~\cite{PandaX:2022ood}.
 
This Letter presents the first results from a blinded analysis of science data from XENONnT, aimed at investigating the nature of the XENON1T excess. Due to its larger active target mass and lower radioactive background, XENONnT is an order of magnitude more sensitive to rare events than its predecessor~\cite{XENON:2020sen}.

The XENONnT experiment~\cite{XENON:2010xwm,XENON100:2011cza,XENON:2017lvq}, located at the INFN Laboratori Nazionali del Gran Sasso\,(LNGS) in Italy, was designed as a fast upgrade of XENON1T~\cite{XENON:2017lvq} and inherits many of its systems such as cooling, gas storage, purification, and Kr-removal~\cite{kr_distillation_xe1t}. At the core of the experiment is a new dual-phase xenon time projection chamber\,(TPC), enclosed in a double-walled stainless-steel\,(SS) cryostat filled with 8.5\,tonnes of liquid xenon\,(LXe). 

The cryostat is suspended at the center of the XENON1T water Cherenkov muon veto~\cite{XENON1T:2014eqx}. A neutron veto was added, surrounding the cryostat, to detect $\gamma\text{-rays}$. produced from neutron capture on Gd, to be added at a later stage. It consists of an
octagonal enclosure with an average diameter of 4\,m and a height of 3\,m
composed of reflective panels and 120 8-inch Hamamatsu R5912 photomultiplier tubes (PMTs). For the first science run, referred to as SR0, the neutron veto was operated with pure water, with an estimated neutron tagging efficiency of $\sim$68\%.

The low-energy ER background in XENON1T was dominated by $^\mathrm{214}$Pb, a $\upbeta$-emitter originating from $^\mathrm{222}$Rn. For XENONnT, in addition to an extensive material radioassay campaign~\cite{XENON:2021mrg}, a new high-flow radon removal system was developed to further reduce this background~\cite{Murra:2022mlr}. The system can operate in two independent modes: the gaseous xenon\,(GXe)-mode, where radon is extracted from warm sections of the detector system (e.g. from around PMT high voltage and signal cables) with a GXe flow of 20\,SLPM before it enters the LXe, and the LXe-mode, where, in addition to the GXe extraction, the entire LXe mass is exchanged every 5.5\,days, which matches the mean lifetime of $^\mathrm{222}$Rn. While commissioning this new system we found that, at the time, its operation in LXe-mode resulted in a drastic drop of xenon purity. For this reason, the system was used only in GXe-mode for SR0, resulting in a $^\mathrm{222}$Rn level of 1.7\,$\mathrm{\upmu}$Bq/kg.

The cylindrical TPC, 1.33\,m in diameter and 1.49\,m tall, encloses an active mass of 5.9\,tonnes of LXe, viewed by two arrays of 3-inch Hamamatsu R11410-21 PMTs~\cite{Antochi:2021pmt}. The top (bottom) array contains 253 (241)\,PMTs, arranged in a hexagonal pattern to maximize the light collection efficiency\,(LCE). The TPC walls are made of PTFE panels.

Five electrodes made of parallel SS wires set the electric fields in the TPC. The anode and gate electrodes, featuring 5\,mm pitch 216\,$\mathrm{\upmu}$m-diameter wires, are positioned 3.0\,mm above and 5.0\,mm below the liquid-gas interface, respectively. Four (two) additional 304\,$\mathrm{\upmu}$m-diameter wires were installed perpendicular to the anode (gate) wires to minimize sagging. The cathode electrode features 304\,$\mathrm{\upmu}$m-diameter wires arranged with a 7.5\,mm pitch. The two remaining electrodes, the top and bottom screens, are positioned 28\,mm above the anode and 55\,mm below the cathode, respectively, and protect the PMTs from high electric fields. 

The cathode and gate electrodes, together with a field cage, define and shape the electric field in the drift region. The field cage consists of two sets of alternating concentric copper rings, connected by two redundant resistive chains. The bottom-most ring is connected to the cathode with a resistor, while the top is independently biased\,(V$_\mathrm{FSR}$). This allows for the tuning of the electric field during operations, improving its homogeneity.

The filling of the TPC was completed in the fall of 2020 and all components were successfully operated for several weeks until November 2020, when a short-circuit between the cathode and bottom screen limited the cathode voltage to $-$2.75\,kV. Therefore, during SR0 the electrodes were kept at V$_\mathrm{anode}$ = +4.9\,kV, V$_\mathrm{gate}$ = +0.3\,kV, V$_\mathrm{cathode}$ = $-$2.75\,kV and V$_\mathrm{FSR}$ = +0.65\,kV, corresponding to a drift field of 23\,V/cm and an electron extraction field of 2.9\,kV/cm in the liquid. With these fields, the response of the TPC was sufficient to acquire science data and exploit the low background level of XENONnT.

Energy depositions above $\sim$1\,keV in LXe can produce detectable prompt scintillation photons\,(S1) at 175\,nm~\cite{FUJII2015293} and ionization electrons which are drifted by the external field and extracted into the GXe above the liquid, where they produce secondary scintillation photons\,(S2) via electroluminescence at the same wavelength. The time difference between the S1 and S2 signals is proportional to the interaction depth $z$.

Both S1 and S2 signals are attenuated due to absorption by impurities dissolved in xenon; thus, high xenon purity is critical for the detector performance. During operation, the xenon is purified continuously using a new cryogenic purification system, based on the work detailed in~\cite{Plante:2022khm}. The LXe is purified through a dedicated adsorbent at a rate of 2 liters per minute, corresponding to 8.3\,tonnes/day. With liquid purification we achieved an electron lifetime, the drift time after which the number of electrons are attenuated to 1/e, larger than 10\,ms, more than an order of magnitude improvement compared to XENON1T. In addition, the GXe volume of the cryostat is continuously purified with a high-temperature getter upstream of the gas inlet to the radon removal system.

Since tritium was a potential explanation for the XENON1T excess, a number of measures were taken to minimize the possibility of introducing it in the form of tritiated hydrogen (HT) and tritiated water (HTO)~\cite{XENON:2020rca}. The TPC was outgassed for a period of about three months before filling the cryostat with GXe. The entire xenon inventory was processed through the Kr-removal system during its transfer into the gas storage system, thereby considerably reducing any initial HT content. In preparation to filling the cryostat with GXe and eventually LXe, the xenon was transferred to the liquid storage system via high temperature getters, which include hydrogen removal units. Prior to cool down and filling, the cryostat and TPC were also treated by continuously circulating GXe for $\sim$3\,weeks. Every time GXe or LXe was filled into the cryostat it was always purified via the getters. Following these measures, the hydrogen removal units were regenerated before the start of SR0.

The SR0 dataset was collected from July 6, 2021 to November 10, 2021 with a total livetime of 97.1\,days. During this period, the detector’s temperature, pressure, and liquid level above the gate electrode remained stable at ($176.8 \pm 0.4$)\,K, ($1.890 \pm 0.004$)\,bar, and ($5.02 \pm 0.20$)\,mm, respectively. The  
TPC PMT gains were set at $\sim$2$\times$10$^{6}$ and were stable within $3\,\%$. The electron extraction efficiency and the mean single electron gain were measured to be $(53\pm3)$\,\% and $(31.2\pm1.0)$\,photoelectrons\,(PE) per extracted electron, respectively.
These values were affected by temporary ramp-downs of the anode, caused by localized, sustained, high-rate bursts of electrons. The effect stabilized within 3 days and is corrected for in the analysis.

The PMT signals are amplified by a factor of 10, digitized at a sampling rate of 100\,MHz, and reconstructed with the open-source software straxen (v1.7.1)~\cite{strax,straxen}. Data is collected in `triggerless' mode; there is no global triggering scheme. Zero length encoding on a per-PMT basis reduces the data volume by only writing regions where a PMT hit crosses its threshold, typically 2.06\,mV corresponding to $\sim$0.3\,PE~\cite{XENON:2019bth}. The mean single-PE acceptance was 91\,\%. 

Out of 494\,PMTs, 17 were either turned off or excluded from analysis due to internal vacuum degradation, light emission, or noise. PMT hits are iteratively grouped with adjacent hits within a 700\,ns time window into clusters, and these clusters are successively subdivided into smaller ones to yield S1 or S2 peaks. Classification of S1s and S2s is performed based
on their hit-pattern and peak shape. S1s require hits from three or more PMTs occurring within 100\,ns (3-fold coincidence). A peak not classified as an S1 with at least four contributing PMTs is considered an S2-peak. An event time-window is defined by an S2 peak above 100\,PE ($\sim$3 extracted electrons) and the preceding maximum drift-time of 2.2\,ms where an S1 may have occurred, extended on both sides by 0.25\,ms. If another S2 above 100\,PE occurs within the duration of the event, the two time-windows are combined.  

The interaction position in the horizontal plane ($r$, $\theta$) of an event is reconstructed using the S2 hit pattern of the top PMT array. Three independent machine learning algorithms (multilayer perceptron, convolutional neural network, and graph constrained network~\cite{posrec:gcn}) are used. The uncertainty of a reconstructed position at the edge of the detector is estimated to be $\sim$1\,cm for an S2 of 1000\,PE, validated with events on the inner surface of the PTFE panel.

Injections of $^\mathrm{83m}$Kr~\cite{kr83m:2009} were performed every two weeks to monitor detector stability and calibrate the energy reconstruction and position dependencies in the detector response. The homogeneity of this internal calibration source is used to correct for the non-uniform drift field, which results in the reconstructed position of events being biased to smaller radii. The drift field as a function of $r$ and $z$ was evaluated by matching the boundary of the $^\mathrm{83m}$Kr distribution to that of a simulation accounting for charge accumulation on the PTFE surface using COMSOL~\cite{comsol, LUX:2017mhm}. This simulation-driven field map was validated using the measured ratio of the two $^\mathrm{83m}$Kr S1 signals~\cite{LUX:2017kr}, and is in agreement with simulation to the level of 3\% in the fiducial volume used in the analysis. The maximum and minimum drift field in the fiducial volume are 21.5\,V/cm and 28.5\,V/cm, respectively.

The S1 signal size is normalized to the mean response in the central part of the TPC to account for spatial dependence of the LCE, determined with $^\mathrm{83m}$Kr data. When evaluating this correction, the effect of the non-uniform drift field on the $^\mathrm{83m}$Kr light yields, as measured in~\cite{Jorg:2021hzu}, up to 1\% inside the fiducial volume, is factored out. This ensures that the correction map only includes geometric effects, so it is applicable to events in a wide energy range. 

The S2 signal size is corrected for the time-dependent electron lifetime, measured with $^\mathrm{83m}$Kr and $^\mathrm{222}$Rn $\upalpha$-decays. The impact of the drift field on the charge yield of the respective source~\cite{Jorg:2021hzu}, up to 5\% for $^\mathrm{83m}$Kr inside the fiducial volume, is considered when determining the electron lifetime. After correction, the measurements with both sources agree, and are consistent with those from a purity monitor installed in the LXe purification system. 

Two additional corrections are applied to S2s. First, the spatially-dependent S2 response due to, e.g., electrode sagging is determined using $^\mathrm{83m}$Kr and used to correct the S2 size based on the reconstructed horizontal position at extraction. No electric field effects were considered for this correction due to the small electric field radial dependence. Second, a time-dependent correction for variations in the single electron gain and extraction efficiency is included to account for transient effects following ramping up of the anode voltage.

In this analysis, we consider ER interactions only. The S1 and S2 response to low energy ERs was calibrated with two radioactive sources. First, $^\mathrm{220}$Rn, whose $\upbeta$-emitting daughter $^\mathrm{212}$Pb produces a uniform ER spectrum at low energies~\cite{XENON:2017rfc}, was injected into the TPC in June 2021.  Second, $^\mathrm{37}$Ar, which primarily produces low-energy depositions of 2.82\,keV via (K-shell) electron capture~\cite{Boulton:2017ar}, was injected in December 2021. Being a mono-energetic and a low-energy source, $^\mathrm{37}$Ar is primarily used to calibrate the response near the energy threshold (and the location of the XENON1T excess). 

Using the $^\mathrm{220}$Rn calibration as a reference, the ER signal region in S1-S2 space was blinded below 20\,keV in the SR0 dataset while the analysis was ongoing. The primary science goal of XENONnT is to search for Weakly Interacting Massive Particles\,(WIMPs), which are expected to produce nuclear recoils (NRs). The NR response was calibrated with an $^\mathrm{241}$AmBe neutron source and used to define the region of interest\,(ROI) for WIMPs, which was kept blinded at the time the analysis for this paper was performed.
 
Like in \cite{XENON:2020rca}, this analysis is performed in the space of reconstructed energy using a combined energy scale defined by $E = W \left( c\mathrm{S1}/g1 + c\mathrm{S2}/g2 \right)$,
where $E$ is the reconstructed energy, the mean energy to produce a quantum, $W$, is fixed to $13.7~\rm eV/quanta$~\cite{dahl2009}, $c$ denotes the corrected S1 and S2 variables, and $g1$ ($g2$) is the gain constant that defines the average S1 (S2) signal in PE observed per unit photon (electron) produced in the interaction. Using mono-energetic calibration sources $^{37}$Ar, $^{83 \mathrm{m}}$Kr,  $^{131 \mathrm{m}}$Xe, and $^{129 \mathrm{m}}$Xe the gain constants are found to be $g1 = (0.151\pm0.001)$\,PE/photon and $g2 = (16.5\pm0.6)$\,PE/electron. These values are extracted after accounting for an energy-dependent `reconstruction bias', a non-linearity in the S1 and S2 reconstruction due to, e.g., single-PE threshold effect and/or PMT afterpulsing. The magnitude of this bias is determined via simulation. This effect, 1\,\% at 41.5\,keV\,($^\mathrm{83m}$Kr) and 2\,\% at 236.1\,keV\,($^\mathrm{129m}$Xe), is modeled empirically in reconstructed energy space and included when constructing the background models. Informed primarily by the $^\mathrm{37}$Ar calibration, the energy resolution is modeled using a skew-Gaussian smearing function~\cite{Szydagis:2021hfh, LUX:2020car, Szydagis:2021ar}, rather than a pure Gaussian as done in ~\cite{XENON:2020rca}.

The ROI for this analysis is (1, 140)\,keV in reconstructed energy, with event-selection criteria similar to~\cite{XENON:2020rca, XENON:2019ykp}. All events must have a valid S1-S2 pair. Signals are expected to induce single scatters; thus, multiple-scatter events are removed. An S2 threshold of 500\,PE is applied. We remove events far below and above the ER region in the cS1-cS2 space to avoid background events originating from $^\mathrm{222}$Rn daughters on the TPC surface~\cite{xe1t_ana_paper_2} and events in GXe, respectively. TPC events occuring within 300\,ns of neutron veto events are removed. 

\begin{figure}[ht]
    \centering
    \includegraphics[width=1\linewidth]{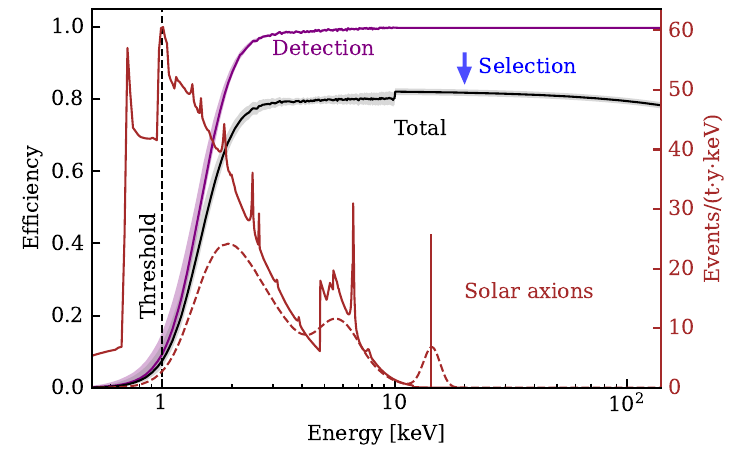}
    \caption{Efficiencies in reconstructed energy and the solar axion signal in true (red solid) and reconstructed energy (red dashed). The purple curve represents the detection efficiency dominated by the 3-fold S1 coincidence requirement. The black curve is the total efficiency, which is a combination of the detection and event-selection efficiencies. The discontinuity at 10\,keV is caused by the still-blinded WIMP search region. The bands indicate the 1$\sigma$ uncertainty. The black dashed line shows the 1\,keV energy threshold of this search. The red solid and dashed lines represent the solar axion signal model before and after accounting for energy smearing and efficiency loss.}
    \label{fig:eff}
\end{figure}

The efficiencies and uncertainties of the event selections are estimated following the procedure in \cite{XENON:2020rca, XENON:2019ykp}. The detection efficiency is dominated by the 3-fold coincidence requirement for valid S1s and was determined using both a data-driven method of sampling PMT hits from S1s using $^\mathrm{37}$Ar and $^\mathrm{83m}$Kr calibration data to mimic low-energy S1s as well as from waveform simulation~\cite{XENON:2019ykp}. The two methods agree within 1\,$\%$, and the waveform simulation method is taken as the nominal one (see Fig.~\ref{fig:eff}). The S2 reconstruction efficiency is determined via simulation to be $\sim$100\,$\%$ for the energies considered here. The combined efficiency of detection and event selection with uncertainties are also shown in Fig.~\ref{fig:eff}. The discontinuity of the combined efficiency at 10\,keV is due to the blinded WIMP search region.

\begin{figure}[ht]
    \centering
    \includegraphics[width=1\linewidth]{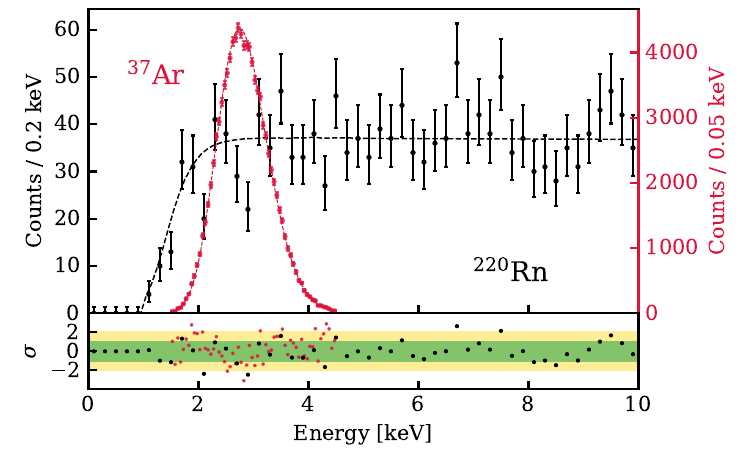}
    \caption{Calibration data and models at low energy. Both $^\mathrm{220}$Rn and $^\mathrm{37}$Ar data are fit using unbinned maximum likelihoods. The $^\mathrm{220}$Rn data fit is performed in the energy interval (1, 140)\,keV, with the low-energy region showing the efficiency near the energy threshold. The $^\mathrm{37}$Ar data validates the energy reconstruction and skew-Gaussian smearing model.}
    \label{fig:ar37_rn220}
\end{figure}

Fig.~\ref{fig:ar37_rn220} shows the reconstructed energy spectra of calibration data at low energy, along with unbinned maximum likelihood fits with the response model used in this analysis. The fit to $^{220}$Rn suggests that the energy threshold, selection efficiency, and energy reconstruction are well-understood. The fit to $^{37}$Ar, which assumes a fixed mean of 2.82\,keV and a fixed resolution, validates the skew-Gaussian smearing model and anchors the energy reconstruction of peaks down to 2.82\,keV.

A fiducial mass of $(4.37 \pm 0.14)$\,tonnes is used for this analysis, yielding a total exposure of 1.16\,tonne$\times$years. The fiducial volume was optimized based on the spatial distribution of the $\gamma\text{-ray}$ induced material background, as well as instrumental backgrounds near the detector wall and liquid-gas interface. The uncertainty of the fiducial mass is dominated by the estimation of the charge-insensitive volume of the TPC, a region near the edge of the detector where extraction of ionization electrons is impossible due to the inhomogeneity in the drift field caused by the limited drift field strength.

We consider three categories of potential BSM signals in this search: a) solar axions, b) solar neutrinos with an enhanced magnetic moment, and c) bosonic DM, which primarily includes Axion-like Particles (ALPs) and dark photon DM. The solar axion model is the one used in \cite{XENON:2020rca} and has been updated to include the Inverse Primakoff effect~\cite{Gao:2020wer, Dent:2020jhf, Abe:2021abe}. The model thus has six components governed by three couplings, i.e. axion couplings to electrons\,($g_\mathrm{ae}$), photons\,($g_\mathrm{a\upgamma}$), and nucleons\,($g_\mathrm{an}^\mathrm{eff}$). An example of the solar axion signal is shown in Fig.~\ref{fig:eff}.

In order to further explore the possibility of tritium as an explanation for the XENON1T excess, we operated XENONnT in a different mode for 14.3\,days bypassing the getter purifying the GXe volume of the cryostat after the SR0 data was collected. Being considerably more volatile than xenon, any outgassed hydrogen (and therefore HT) is more effectively removed by this purification scheme. Furthermore, the shorter purification time constant\,($\sim$2\,h), compared to that of liquid purification\,($\sim$1\,day), also makes the removal more effective. Therefore, the equilibrium concentration of HT in LXe is expected to be enhanced by bypassing this getter. The enhancement factor is conservatively estimated to be at least 10, but could be as large as 100. This `tritium-enhanced' dataset, when unblinded, showed no evidence for a tritium-like excess. Based on this null result, combined with the aforementioned reduction measures, tritium is not included in the background model.

$^{37}$Ar was suggested as another potential source of the XENON1T excess~\cite{Szydagis:2021ar}. This hypothesis was ruled out due to the strong constraints set on trace amounts of $^{37}$Ar both from cosmogenic activation and from a potential air leak~\cite{XENON:2020rca}. Similar considerations apply to XENONnT. As mentioned earlier, the entire xenon inventory was cryogenically distilled underground by the Kr-removal system, which is also extremely effective in reducing $^{37}$Ar to a negligible level~\cite{XENON:2021fkt}. The variation of $^\mathrm{nat}$Kr concentration in xenon over the SR0 period, measured with rare-gas mass spectrometry~\cite{Lindemann:2014kr}, sets an upper limit of $4\times10^{-6}$\,mbar$\cdot$l/s on any air leak that leads to a negligible level of $^{37}$Ar. For this reason, $^{37}$Ar is also not included in the background model.

\begin{table}[ht]
	\centering
	\caption{The background model $B_0$ with fit constraints and best-fit number of events for each component in (1, 140)\,keV.}
	\begin{tabular}{
			>{\centering}m{2.8cm} 
			>{\raggedleft}m{1.3cm}
			c
			>{\raggedright}m{1.3cm} 
			>{\raggedleft}m{0.8cm} 
			c 
			>{\raggedright\arraybackslash}m{0.8cm}
		}
		\hline\hline
		Component  & \multicolumn{3}{c}{Constraint} & \multicolumn{3}{c}{Fit}   \\
		
		\hline
		$^\mathrm{214}$Pb & \multicolumn{3}{c}{(570, 1200)} &  960 &$\pm$& 120\\
		$^\mathrm{85}$Kr & 90 & $\pm$ & 60  & 90 & $\pm$ & 60\\
		Materials & 270 &$\pm$& 50 & 270 & $\pm$ & 50 \\
		$^\mathrm{136}$Xe & 1560 &  $\pm$ & 60 & 1550 &  $\pm$ & 50 \\
		Solar neutrino & 300 & $\pm$ & 30 & 300 & $\pm$ & 30 \\
		$^\mathrm{124}$Xe & \multicolumn{3}{c}{-} & 250 & $\pm$ & 30 \\
		AC & 0.70 & $\pm$ & 0.04 & 0.71 & $\pm$ & 0.03 \\
		$^\mathrm{133}$Xe & \multicolumn{3}{c}{-} & 150 & $\pm$ & 60\\
		$^\mathrm{83m}$Kr & \multicolumn{3}{c}{-} & 80 & $\pm$ & 16 \\
		\hline\hline
	\end{tabular}
	\label{tab:backgrounds}
\end{table}

We consider nine components in the background model $B_0$, as listed in Tab.~\ref{tab:backgrounds}. The dominant background at low energies is still the $\upbeta$-decay of $^\mathrm{214}$Pb, the activity of which is bound between ($0.777\pm0.006_\mathrm{stat}\pm0.032_\mathrm{sys}$)\,$\mathrm{\upmu}$Bq/kg and ($1.691\pm0.006_\mathrm{stat}\pm0.072_\mathrm{sys}$)\,$\mathrm{\upmu}$Bq/kg. These bounds are determined by the rates of $^\mathrm{218}$Po and $^\mathrm{214}$Po $\upalpha$-decays, parent and daughter of $^\mathrm{214}$Pb in the decay chain, respectively. We used the $^\mathrm{214}$Pb spectrum from \cite{Haselschwardt:2020iey} that is calculated as a forbidden transition. The $^{220}$Rn rate is less than 5\% of the $^{222}$Rn rate and thus the $\upbeta$-decay of $^{212}$Pb is not included in the background model.

The $^\mathrm{nat}$Kr concentration was measured to be ($56\pm36$)\,ppq. The abundance of $^\mathrm{85}$Kr in $^\mathrm{nat}$Kr is taken as $2\times10^{-11}$ based on seasonal measurements of $^\mathrm{85}$Kr activity in the LNGS air of 1.4\,$\mathrm{Bq}/\mathrm{m^3}$, which is consistent with \cite{BOLLHOFER20197}. Those measurements are then propagated to constrain the $^{85}$Kr rate with an uncertainty of $\sim$66\,\%, dominated by the $^\mathrm{nat}$Kr concentration measurement.

Gamma-ray backgrounds from materials are found to have a flat spectrum below 140\,keV in the fiducial volume from Geant4 simulation~\cite{GEANT4:2002zbu, Allison:2016lfl}. We expect the rate to be ($2.1\pm0.4$)\,events/(tonne$\times$year$\times$keV) (abbreviated as events/(t$\cdot$y$\cdot$keV) for the rest of the paper), where the uncertainty originates from the simulation and from the measurement uncertainties of material radioassay~\cite{XENON:2021mrg}.

Having successfully reduced other sources, the two-neutrino double-beta ($2\upnu\upbeta\upbeta$) decay of $^\mathrm{136}$Xe becomes an important background in this analysis, overtaking $^\mathrm{214}$Pb as the dominant component above 40\,keV. This background is constrained by an in-situ measurement of the xenon isotopic abundance with a residual gas analyzer and the half-life from \cite{EXO-200:2013xfn}. We also allow for a shape change to account for the uncertainty on the theoretical calculation of this spectrum, $\sim$1.5\% in our ROI and specifically whether this isotope is better described by the higher state dominance~\cite{Kotila:2012zza} or single state dominance~\cite{Kotila:2020lower} model of $2\upnu\upbeta\upbeta$ decay.

The double-electron capture (2$\upnu$ECEC) decay rate of $^\mathrm{124}$Xe is left unconstrained in $B_0$. The energy spectrum adopts the updated model of \cite{XENON:2022met}, which takes into account the contributions from higher atomic shells compared to \cite{XENON:2020rca} and uses fixed branching ratios. The reconstruction of the dominant KK-capture peak at 64.3\,keV was also used as validation of the energy reconstruction.

The spectrum of electron scattering from solar neutrinos is computed as in~\cite{XENON:2020rca}. We assign a 10\,\% solar neutrino flux uncertainty based on the Borexino measurement~\cite{BOREXINO:2018ohr}. $^\mathrm{133}$Xe was produced by neutron activation from the $^{241}$AmBe calibration several months before the SR0 science data taking and a tiny fraction survived to the start of SR0. Given that it does not impact the low-energy region and this rate is small, the background is allowed to vary freely in the fit. Trace amounts of $^\mathrm{83m}$Kr leftover from calibrations are also present in the SR0 data, the rate of which is also left unconstrained. 

The last background component, accidental coincidences (ACs), is the only non-ER background in $B_0$. Uncorrelated S1s and S2s can randomly pair and form fake events, and a small fraction survives all event selections~\cite{XENON:2019ykp}. AC events overlap with the ER band in cS1-cS2 space and produce a spectrum that increases towards low energies. Its rate in the ER region is predicted to be $(0.61\pm0.03)$\,events/(t$\cdot$y) using a data-driven method, which randomly pairs isolated S1s and isolated S2s data into fake events and subsequently applies the aforementioned event selections.

 \begin{figure}[ht]
    \centering
    \includegraphics[width=1\linewidth]{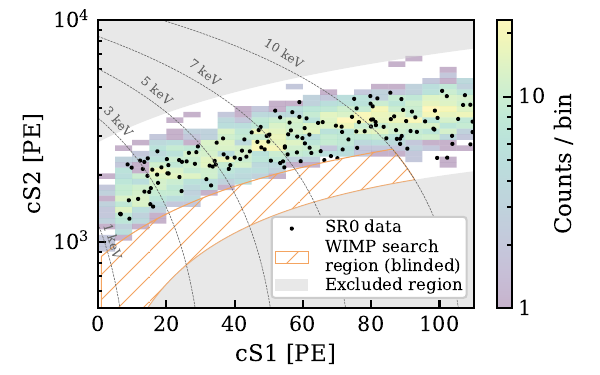}
    \caption{Science data\,(black dots) in the cS1-cS2 space, overlaid on $^{220}$Rn data\,(2D histogram). The WIMP search region\,(orange) is still blinded and not used in this search. Regions\,(gray shaded) far away from the ER band are excluded to avoid anomalous backgrounds. Iso-energy lines are represented by the gray dashed lines.}
    \label{fig:s1s2}
\end{figure}

After all aspects of the analysis had been fixed and a good agreement between the background model and data above 20\,keV was found (p-value $\sim$0.2), the region between $\pm 2\sigma$ quantile of ER events in S2 was unblinded. The NR region below ER $-2\,\sigma$ remains blinded while the WIMP analysis continues, as shown in Fig.~\ref{fig:s1s2}.

\begin{figure}[ht]
    \centering
    \includegraphics[width=1\linewidth]{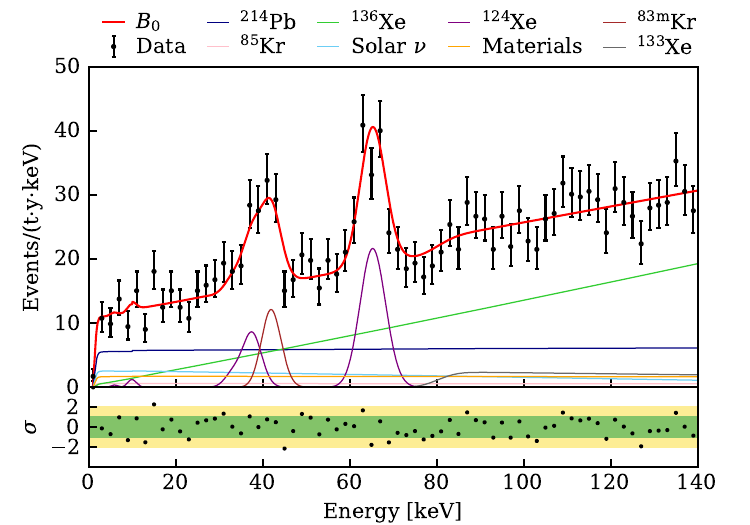}
    \caption{Fit to SR0 data using the background model $B_0$. The fit result of $B_0$ is the red line. The subdominant AC background is not shown.}
    \label{fig:bkg}
\end{figure}

\begin{figure}[ht]
    \centering
    \includegraphics[width=1\linewidth]{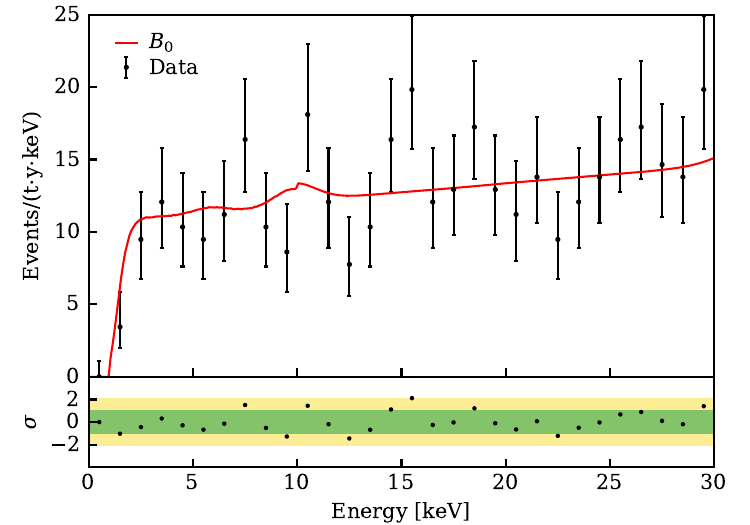}
    \caption{Data and best-fit $B_0$ model below 30\,keV. No significant excess above the background was found. The bump at $\sim$10\,keV is from the LL-shell of $^\mathrm{124}$Xe 2$\upnu$ECEC~\cite{XENON:2022met}, while the discontinuity at 10\,keV is caused by the blinded WIMP search region, see Fig.~\ref{fig:eff} and \ref{fig:s1s2}. A finer binning than in Fig.~\ref{fig:bkg} is used to show the event rate change near the threshold.}
    \label{fig:lower}
\end{figure}

We performed a fit in reconstructed energy space using an unbinned maximum likelihood similar to that in \cite{XENON:2020rca}. The efficiency at low energies is allowed to vary within its uncertainty band. The best-fit of $B_0$ is illustrated in Fig.~\ref{fig:bkg} and Fig.~\ref{fig:lower}, and the results are listed in Tab.~\ref{tab:backgrounds}. The SR0 dataset agrees well with $B_0$, and no excess above the background is found. The efficiency-corrected average ER background rate within (1, 30)\,keV is measured to be ($15.8\pm1.3_\mathrm{stat}$)\,events/(t$\cdot$y$\cdot$keV), a factor of $\sim$5 lower than the rate in XENON1T~\cite{XENON:2020rca}. This is the lowest background rate ever achieved at these energies among dark matter direct detection experiments. The spectral shape in Fig.~\ref{fig:bkg} is, for the first time, mostly determined by two second-order weak processes: the $2\upnu\upbeta\upbeta$ of $^\mathrm{136}$Xe and 2$\upnu$ECEC of $^\mathrm{124}$Xe.

The best-fit activity concentration of $^\mathrm{214}$Pb is ($1.31\pm 0.17_\mathrm{stat}$)\,$\mathrm{\upmu}$Bq/kg assuming the branching ratio to the ground state is 12.7\,\%~\cite{Zhu:2021qss}. The best-fit rate of $^{124}$Xe 2$\upnu$ECEC translates to a half-life of $T_{1/2}^\mathrm{2\upnu ECEC} = (1.18\pm0.13_\mathrm{stat}\pm0.14_\mathrm{sys})\times10^{22}\,\mathrm{yr}$, where the 12\,\% systematic uncertainty is from selection efficiency\,(8\,\%), exposure\,(3\,\%), $^{124}$Xe abundance\,(6\,\%), and capture fraction\,(6\,\%). This result is consistent with the half-life reported in~\cite{XENON:2022met}.

\begin{figure*}[ht]
    \centering
    \includegraphics[width=\textwidth]{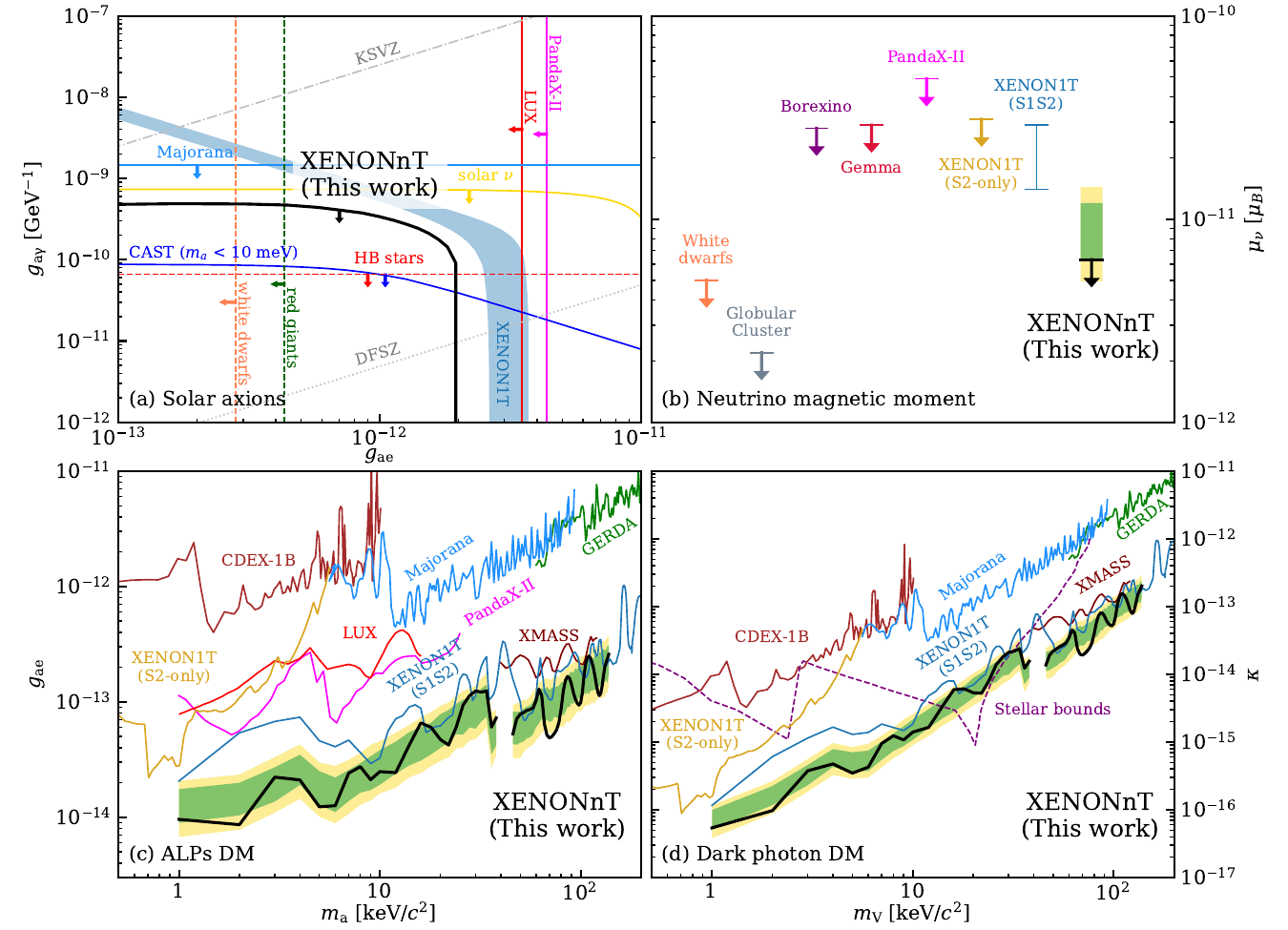}
    \caption{90\% C.L. upper limit on different new physics models. Constraints on the axion-electron $g_{\mathrm{ae}}$ and axion-photon $g_\mathrm{a\upgamma}$ couplings from a search for solar axions are shown in (a). Constraints on solar neutrinos with an enhanced magnetic moment\,(b), ALP DM\,(c), and dark photon DM\,(d) are shown together with the 1\,$\sigma$ (green) and 2\,$\sigma$ (yellow) sensitivity bands estimated with the background-only fit. Constraints between (39, 44)\,keV/c$^2$ are excluded in (c) and (d) due to the unconstrained $^\mathrm{83m}$Kr background. Selected limits from other experiments~\cite{CAST_gae,borexino_nmm, Gemma_mu, CDEX:2019exx,PandaX:2021nmm,EDELWEISS:2018tde, GERDA:2020emj, LUX:2017glr, Majorana:2016hop, PandaX:2017ock, SuperCDMS:2019jxx, XENON100:2017pfn, XENON:2019gfn, XMASS:2018pvs, Majorana:2022bse} and astrophysical observations~\cite{revisiting_axionphoton_hb,redgiant_gae,wdlf_limit,Corsico_2014, An:2014twa} are also shown.
    }
    \label{fig:limits}
\end{figure*}

Fig.~\ref{fig:limits} shows the 90\,$\%$ confidence level (C.L.) upper limit on solar axions, bosonic DM models, and solar neutrinos with an enhanced magnetic moment together with sensitivity bands estimated from the background-only fit. The solar axion limit is the upper bound of a 3D confidence volume evaluated in the space of $g_\mathrm{ae}$, $g_\mathrm{a\upgamma}$, and $g_\mathrm{an}^\mathrm{eff}$, and projected onto the $g_\mathrm{a\upgamma}$ vs. $g_\mathrm{ae}$ space. The 90\,\% C.L. upper limit on the 14.4\,keV peak from the solar axion $^{57}$Fe component is 20\,events/(t$\cdot$y). Those limits hold for axion masses up to $\sim$100\,eV/$c^2$. Since the $^{83\mathrm{m}}$Kr rate is left unconstrained, we only place upper limits for bosonic DM with a mass between (1, 39) and (44, 140)\,keV/c$^2$. The maximum local significance of this search is around 1.8\,$\sigma$. The excess observed in the XENON1T experiment~\cite{XENON:2020rca}, when modeled as a 2.3\,keV mono-energetic peak, is excluded with a statistical significance of $\sim$4$\,\sigma$. The 90\,$\%$ C.L. upper limit on solar neutrinos with an enhanced magnetic moment is $\mu_{\upnu} < 6.4 \times 10^{-12}\,\mu_\mathrm{B}$.

We also searched for a tritium component on top of the background model $B_0$. The best-fit rate of tritium is 0 and the upper limit\,(90\,\% C.L.) is 15\,events/(t$\cdot$y), corresponding to a concentration of $5.8\times10^{-26}$\,mol/mol of tritium in xenon. If the excess observed in XENON1T was from trace amounts of tritium, the disappearance of the excess in XENONnT may have resulted from the aforementioned rigorous tritium prevention measures. 

In summary, we performed a search for new physics in the electronic recoil data in the keV energy range from XENONnT using an exposure of 1.16\,tonne$\times$years. The average ER background rate of ($15.8\pm1.3_\mathrm{stat}$)\,events/(t$\cdot$y$\cdot$keV) in the (1, 30)\,keV energy region is the lowest ever achieved in a DM search experiment. The blind analysis shows no excess above the background, excluding our previous BSM interpretations of the XENON1T excess. Upper limits on solar axions, bosonic DM, and solar neutrinos with an enhanced magnetic moment are set, excluding new parameter spaces. A measured half-life of $^\mathrm{124}$Xe 2$\upnu$ECEC is also reported, consistent with the final XENON1T measurement~\cite{XENON:2022met}. 

XENONnT is continuing to take data at LNGS. Since the conclusion of SR0, upgrades to the radon removal system were made to allow for its operation in the originally foreseen combined LXe and GXe mode, resulting in a further reduction of the $^{222}$Rn activity concentration $< 1\,\mathrm{\upmu}\mathrm{Bq/kg}$. This extremely low background level coupled with a large target mass will allow XENONnT to continue probing intriguing physics channels such as DM, solar axions, and solar neutrinos well into the future. 

We thank Dr.~Jenni Kotila at University of Jyv\"askyl\"a for providing the single state dominance spectrum of $^{136}$Xe $2\upnu\upbeta\upbeta$ decay and suggesting the treatment for its spectral uncertainty. We also thank Dr.~Roland Purtschert at Climate and Environmental Physics, University of Bern, for measurements of the $^{37}$Ar concentration at LNGS. We gratefully acknowledge support from the National Science Foundation, Swiss National Science Foundation, German Ministry for Education and Research, Max Planck Gesellschaft, Deutsche Forschungsgemeinschaft, Helmholtz Association, Dutch Research Council (NWO), Weizmann Institute of Science, Israeli Science Foundation, Binational Science Foundation, Funda\c{c}\~ao para a Ci\^encia e a Tecnologia, R\'egion des Pays de la Loire, Knut and Alice Wallenberg Foundation, Kavli Foundation, JSPS Kakenhi and JST FOREST Program in Japan, Tsinghua University Initiative Scientific Research Program and Istituto Nazionale di Fisica Nucleare. This project has received funding/support from the European Union’s Horizon 2020 research and innovation programme under the Marie Sk\l{}odowska-Curie grant agreement No 860881-HIDDeN. Data processing is performed using infrastructures from the Open Science Grid, the European Grid Initiative and the Dutch national e-infrastructure with the support of SURF Cooperative. We are grateful to Laboratori Nazionali del Gran Sasso for hosting and supporting the XENON project.

\bibliography{bibliography}

\end{document}